\documentclass[a4paper,fleqn]{cas-sc}
\usepackage[utf8]{inputenc} 
\usepackage{xcolor}
\usepackage{hyperref}
\hypersetup{
  colorlinks=true,
  linkcolor=blue,
  filecolor=magenta,
  urlcolor=cyan,
  citecolor=cyan,
}
\usepackage{graphicx}

\begin{document}

\shorttitle{Quantum Realization of the Wallis Formula}
\shortauthors{Yin Lei et~al.}

\title[mode = title]{Quantum Realization of the Wallis Formula}

\author[1]{Ye Bin}[style=chinese]
\author[1]{Chen Ruitao}[style=chinese]
\author[1]{Yin Lei}[%
   style=chinese,
   orcid= 0000-0002-8585-9759
   ] 
\cormark[1]
\ead{lei@scnu.edu.cn}

\affiliation[1]{organization={School of Materials and New Energy, South China Normal University},
                city={Shanwei},
                postcode={516625}, 
                country={China}
                }
                

\cortext[cor1]{Corresponding author}

\begin{abstract} 
We present a unified quantum-mechanical derivation of the Wallis formula from two solvable radial systems: the circular states of the three-dimensional isotropic harmonic oscillator and the lowest-radial-branch states of the planar Fock--Darwin problem, including the lowest Landau level sector. In both cases, the radial probability density has the exact form $P(r)\propto r^\nu e^{-\lambda r^2}$, which yields the scale-independent reciprocal observable $Q=\langle r\rangle\langle r^{-1}\rangle$. The two systems realize the even and odd half-integer Gamma-function branches of the same moment formula, so that the associated finite Wallis partial products are determined by $Q$ in one case and by $Q^{-1}$ in the other. In the large-angular-momentum regime, the corresponding states become localized on a thin spherical shell or a narrow annulus, with vanishing relative radial width, so that $Q\to1$ and both finite-product representations reduce to the Wallis formula for $\pi$.
\end{abstract}

\begin{keywords}
Wallis formula, semiclassical limit, harmonic oscillator, Fock-Darwin states
\end{keywords}

\maketitle

\section{Introduction}

The Wallis product
$$
\frac{\pi}{2}
=
\prod_{n=1}^{\infty}
\frac{(2n)^2}{(2n-1)(2n+1)}
$$
is a classical identity closely connected with Beta and Gamma functions, trigonometric integrals, and asymptotic analysis. From a quantum-mechanical perspective, the interest of the Wallis formula lies not only in the reappearance of a classical analytic identity, but also in identifying the simple physical structure that supports it. In this sense, the problem is not merely to reproduce $\pi$ in another setting, but to understand how a familiar mathematical formula can emerge from an exact and physically transparent quantum mechanism.

A striking connection between the Wallis product and quantum mechanics was first exhibited by Friedmann and Hagen \cite{FriedmannHagen2015}
through a variational treatment of the hydrogen atom in the large-angular-momentum regime. Later work clarified that the relevant Wallis structure is not uniquely tied to a particular trial function \cite{ChashchinaSilagadze2017}, and that the phenomenon is not restricted to the Coulomb problem alone \cite{CorteseGarcia2018}.

The present formulation differs from these earlier routes in two respects. First, it does not rely on a variational comparison between approximate and exact energies, but instead works directly with exact radial probability densities and the associated reciprocal observable. Second, it does not proceed from one model to another by duality. Rather, it identifies a single radial-Gaussian mechanism realized exactly in two canonical branches, one three-dimensional and one planar. In this way, the emphasis is shifted from model-specific derivations of the Wallis formula to a common exact structure that organizes both realizations.

The present paper addresses this issue in two standard and physically distinct quantum systems: the circular states of the three-dimensional isotropic harmonic oscillator and the lowest-radial-branch states of the planar Fock--Darwin problem, including the lowest Landau level sector. Although their physical interpretations differ, both possess radial probability densities of the form:
$$
P(r)\propto r^\nu e^{-\lambda r^2},
$$
and it is this shared radial-Gaussian structure that organizes the analysis.

The central quantity in our treatment is the dimensionless reciprocal radial observable
$$
Q:=\langle r\rangle \langle r^{-1}\rangle .
$$
For a perfectly sharp classical circular orbit of radius $r_c$, one has $r_c \cdot r_c^{-1}=1$. For a quantum radial distribution, by contrast, $Q\ge 1$,
so that $Q-1$ measures the departure from exact reciprocal radial rigidity. In the large-angular-momentum regime, both systems develop increasingly sharp circular localization: in the oscillator case, a thin spherical shell around the classical orbit; in the planar magnetic case, a narrow annulus with a natural guiding-center interpretation. In both settings the absolute radial width remains finite while the relative width decreases, and accordingly $Q\to 1$. In this sense, the approach $Q\to 1$ may be viewed as a quantitative signature of the correspondence-principle regime, in which the quantum radial distribution becomes effectively concentrated on a classical circular orbit or annulus.

The novelty of the present work is therefore not the general observation that Wallis-type products may arise in quantum mechanics, but the identification of a single radial-Gaussian reciprocal mechanism realized exactly in two canonical solvable models. Analytically, the two systems occupy complementary half-integer Gamma-function branches; physically, they provide two distinct semiclassical realizations of the same tendency toward circular radial rigidity. The Wallis product then appears as the finite-product expression of this common reciprocal-radial structure, rather than as an accidental by-product of unrelated examples.

The paper is organized as follows. Section 2 presents the common radial-Gaussian framework and derives the exact scale-free moment ratio governing $Q$. Section 3 treats the circular states of the three-dimensional isotropic harmonic oscillator, while Section 4 analyzes the lowest-radial-branch Fock--Darwin states. Section 5 summarizes the shared mechanism and its semiclassical meaning.

\section{Common radial-Gaussian framework}
\label{sec:comm-radi-gauss}

The two models studied below share the same radial structure. Their probability densities belong to the family
\begin{align}
P_{\nu,\lambda}(r)=N_{\nu,\lambda} \, r^\nu e^{-\lambda r^2},
\quad r>0,\; \nu>-1,\; \lambda>0,
\label{eq:1}
\end{align}
 with normalization
$$
\int_0^\infty P_{\nu,\lambda}(r) \,\mathrm{d}r=1 \;.
$$
Here, $\nu$ characterizes the shape of the distribution, while $\lambda$ sets its radial scale.

Using the standard Gamma integral
$$
\int_0^\infty r^{a-1}e^{-\lambda r^2} \,\mathrm{d}r
=
\frac{1}{2}\lambda^{-a/2}\Gamma\!\left(\frac{a}{2}\right),
\qquad a>0,
$$
one finds
\begin{equation}
  \begin{aligned}
N_{\nu,\lambda}
=
\frac{2\lambda^{(\nu+1)/2}}{\Gamma\!\left(\frac{\nu+1}{2}\right)},
\label{eq:2}
\end{aligned}
\end{equation}
so that
\begin{equation}
  \begin{aligned}
P_{\nu,\lambda}(r)
=
\frac{2\lambda^{(\nu+1)/2}}{\Gamma\!\left(\frac{\nu+1}{2}\right)} \, r^\nu e^{-\lambda r^2}.
\label{eq:3}
\end{aligned}
\end{equation}
Accordingly, for any real $k$ such that $\nu+k>-1$,
\begin{equation}
  \begin{aligned}
\langle r^k\rangle_{\nu,\lambda}
=
\int_0^\infty r^k P_{\nu,\lambda}(r) \,\mathrm{d}r
=
\lambda^{-k/2}
\frac{\Gamma\!\left(\frac{\nu+k+1}{2}\right)}
{\Gamma\!\left(\frac{\nu+1}{2}\right)}.
\label{eq:4}
\end{aligned}
\end{equation}
We now introduce the dimensionless reciprocal observable
\begin{equation}
  \begin{aligned}
    Q_\nu:=\langle r\rangle_{\nu,\lambda}\langle r^{-1}\rangle_{\nu,\lambda}.
    \label{eq:5}
\end{aligned}
\end{equation}
Substituting the moment formulas gives
\begin{equation}
  \begin{aligned}
Q_\nu
=
\frac{\Gamma\!\left(\frac{\nu+2}{2}\right)\Gamma\!\left(\frac{\nu}{2}\right)}
{\Gamma\!\left(\frac{\nu+1}{2}\right)^2}.
\label{eq:6}
\end{aligned}
\end{equation}


The scale parameter $\lambda$ cancels identically, so $Q_\nu$ depends only on the shape of the radial distribution.

A universal lower bound follows immediately. Since $r>0$, the Cauchy-Schwarz inequality applied to $f=r^{1/2}$ and $g=r^{-1/2}$ yields
\begin{equation}
\begin{aligned}
\biggl(\int_0^\infty f(r)g(r)P_{\nu,\lambda}(r),dr\biggr)^2
\le 
\biggl(\int_0^\infty f(r)^2P_{\nu,\lambda}(r),dr\biggr) 
\biggl(\int_0^\infty g(r)^2P_{\nu,\lambda}(r),dr\biggr),
\end{aligned}
\end{equation}
hence
\begin{equation}
  \begin{aligned}
    1\le \langle r\rangle_{\nu,\lambda}\langle r^{-1}\rangle_{\nu,\lambda}=Q_\nu.
 \label{eq:7}
\end{aligned}
\end{equation}
Thus $Q_\nu-1$ measures the departure from exact reciprocal radial rigidity. In the large-angular-momentum regimes considered later, this excess becomes small because the radial distribution becomes relatively sharp around a classical circular structure.

The Wallis product enters when $Q_\nu$ is evaluated on the half-integer Gamma branches selected by the two quantum systems, for the oscillator circular states, 
$$
\nu=2\ell+2,
$$
whereas for the lowest-radial-branch Fock-Darwin/Landau states,
$$
\nu=2m+1.
$$
These correspond respectively to the even and odd half-integer channels of the same exact Gamma-ratio formula. The finite Wallis products obtained in Sections 3 and 4 are therefore not independent accidents, but two realizations of one common radial-Gaussian reciprocal structure.

\section{Three-dimensional isotropic harmonic oscillator}

Consider the three-dimensional isotropic harmonic oscillator
\begin{equation}
  \begin{aligned}
H=\frac{p^2}{2M}+\frac12 M\omega^2 r^2 .
\end{aligned}
\end{equation}
Its stationary states separate in spherical coordinates as
$$
\psi_{n_r\ell m}(r,\theta,\phi)=R_{n_r\ell}(r)Y_{\ell m}(\theta,\phi),
$$
with energies
\begin{equation}
  \begin{aligned}
E_{n_r,\ell}=\hbar\omega\left(2n_r+\ell+\frac32\right).
\end{aligned}
\end{equation}
For fixed total excitation $N=2n_r+\ell,$ the states with maximal angular momentum are those with $n_r=0$. These circular states have no radial nodes and provide the natural oscillator realization of circular motion. With $\alpha:=\frac{M\omega}{\hbar}$ , their radial wave function is
\begin{equation}
  \begin{aligned}
R_{0\ell}(r)=N_\ell\, r^\ell e^{-\alpha r^2/2},
\end{aligned}
\end{equation}
so the normalized radial probability density ~(\ref{eq:3}) becomes
\begin{equation}
  \begin{aligned}
    P_\ell(r)=|R_{0\ell}(r)|^2r^2 = \frac{2\alpha^{\ell+3/2}}{\Gamma\!\left(\ell+\frac32\right)} \,r^{2\ell+2}e^{-\alpha r^2}.
    \label{eq:8}
\end{aligned}
\end{equation}
Thus the oscillator circular family realizes the common radial-Gaussian class with
$\nu=2\ell+2,\; \lambda=\alpha$, that is, the even branch of the general framework. The extra factor $r^2$ comes from the three-dimensional radial measure, so the branch selection is fixed directly by the geometry of the problem.

Using the moment formula ~(\ref{eq:4}) of Section \ref{sec:comm-radi-gauss}, one obtains
\begin{equation}
  \begin{aligned}
    \langle r\rangle_\ell = \alpha^{-1/2} \frac{\Gamma(\ell+2)}{\Gamma\!\left(\ell+\frac32\right)}, \quad \langle r^{-1}\rangle_\ell = \alpha^{1/2} \frac{\Gamma(\ell+1)}{\Gamma\!\left(\ell+\frac32\right)},
    \label{eq:9}
\end{aligned}
\end{equation}
and hence the dimensionless $Q$ (\ref{eq:6}) becomes 
\begin{equation}
  \begin{aligned}
    Q_\ell^{(\mathrm{osc})} := \langle r\rangle_\ell\langle r^{-1}\rangle_\ell = \frac{\Gamma(\ell+1)\Gamma(\ell+2)} {\Gamma\!\left(\ell+\frac32\right)^2}.
    \label{eq:10}
\end{aligned}
\end{equation}

Define the finite Wallis partial product
\begin{equation}
  \begin{aligned}
W_n:=\prod_{k=1}^{n}\frac{(2k)^2}{(2k-1)(2k+1)} \;,
\label{eq:11}
\end{aligned}
\end{equation}
then the half-integer Gamma identities yield
\begin{equation}
  \begin{aligned}
W_{\ell+1}=\frac{\pi}{2}\,Q_\ell^{(\mathrm{osc})}.
\label{eq:12}
\end{aligned}
\end{equation}
In this form, the exact reciprocal observable determines the finite Wallis product associated with the oscillator branch.

The semiclassical interpretation is equally direct. The reduced radial equation for $u(r)=rR(r)$ takes the one-dimensional form
$$
-\frac{\hbar^2}{2M}u''(r)+V_{\mathrm{eff}}(r)u(r)=Eu(r),
$$
with effective potential
$$
V_{\mathrm{eff}}(r) = \frac{\hbar^2\ell(\ell+1)}{2Mr^2} +\frac12 M\omega^2 r^2.
$$
Its minimum gives the stable classical circular radius
$$
\frac{\mathrm{d}V_{\mathrm{eff}}}{\mathrm{d}r}=0
\quad\Longrightarrow\quad
r_c^2=\frac{1}{\alpha}\sqrt{\ell(\ell+1)}.
$$
On the quantum side, the radial density $P_\ell(r)$ reaches its maximum at
$$
\frac{\mathrm{d}}{\mathrm{d}r}\log P_\ell(r)=0
\quad \Longrightarrow \quad
r_*^2=\frac{\ell+1}{\alpha}.
$$
Hence
$$
\frac{r_*}{r_c}\to 1, \quad  \text{as} \quad \ell\to\infty.
$$
Moreover, expanding $\log P_\ell(r)$ around $r=r_*$ gives
\begin{equation}
  \begin{aligned}
    P_\ell(r)\approx P_\ell(r_*)\exp\!\left[-\frac{(r-r_*)^2}{2\sigma_r^2}\right], \quad \sigma_r \equiv \frac{1}{2\sqrt{\alpha}},
    \label{eq:13}
\end{aligned}
\end{equation}
so that
\begin{equation}
  \begin{aligned}
\frac{\sigma_r}{r_*} = \frac{1}{2\sqrt{\ell+1}} \sim \frac12\,\ell^{-1/2}.
\label{eq:14}
\end{aligned}
\end{equation}
The large$-\ell$ circular states therefore form a thin spherical shell of increasing radius and asymptotically vanishing relative thickness. This is the relevant correspondence-principle regime: the quantum radial distribution becomes concentrated on the classical circular orbit.

Accordingly,
\begin{align}
Q_\ell^{(\mathrm{osc})} = 1+\frac{1}{4\ell}+O(\ell^{-2}), \qquad \ell\to\infty,
  \label{eq:15}
\end{align}
by the standard asymptotic expansion of Gamma-function ratios \footnote{A standard large-$z$ formula is $$
  \frac{\Gamma(z+a)}{\Gamma(z+b)}=  z^{a-b}  \bigg[  1+\frac{(a-b)(a+b-1)}{2z} +O(z^{-2})  \bigg], \quad z\to\infty
  $$.} \cite{Olver2010NIST,tricomi1951asymptotic,DLMFGamma}. Therefore
\begin{align}
W_{\ell+1} = \frac{\pi}{2}\,Q_\ell^{(\mathrm{osc})} \longrightarrow \frac{\pi}{2}, \qquad \ell\to\infty.
  \label{eq:16}
\end{align}
Thus the Wallis formula arises here because the oscillator circular states furnish an exact finite-product realization whose reciprocal-radial observable tends to its classical value in the semiclassical limit.


\section{Lowest-radial-branch Fock--Darwin states}

Consider a charged particle of mass $\mu$ moving in the plane under a uniform magnetic field $B$, together with an isotropic harmonic confinement of frequency $\omega_0$:
$$
H=\frac{1}{2\mu}(\mathbf p+e\mathbf A)^2+\frac12 \mu\omega_0^2 r^2.
$$
In the symmetric gauge

\begin{align}
  \mathbf A=\left(-\frac{By}{2},\frac{Bx}{2},0\right),
  \label{eq:17}
\end{align}
expanding the minimal-coupling Hamiltonian gives the Fock-Darwin form \cite{Fock1928,Darwin1931} :
$$
H=\frac{\mathbf p^2}{2\mu}+\frac12 \mu\Omega^2 r^2+\frac{\omega_c}{2}L_z \;, 
$$
with $\omega_c=\frac{eB}{\mu}, \; \Omega=\sqrt{\omega_0^2+\frac{\omega_c^2}{4}}$.
The symmetric gauge (\ref{eq:17}) is adopted here for convenience, since it makes the rotationally adapted radial-Gaussian structure explicit; the probability density and the observable $Q$ are attached to the physical state rather than to the gauge choice itself.

We consider the lowest-radial-branch states $n_r=0$ with $m\ge 0$, which include the lowest-Landau-level sector in the limit $\omega_0\to 0$~\cite{Landau1930}. Their wavefunctions may be written as
\begin{equation}
\psi_{0m}(r,\theta)=N_m\,r^m e^{-\beta r^2/2}e^{im\theta}, \; \beta:=\frac{\mu\Omega}{\hbar} \;,
\label{eq:18}
\end{equation}
with the normalization constant $N_m=\sqrt{\frac{\beta^{m+1}}{\pi\,m!}}$ , and the normalized radial probability density is
\begin{align}
P_m(r)=2\pi |\psi_{0m}(r,\theta)|^2\,r = \frac{2\beta^{m+1}}{m!}\,r^{2m+1}e^{-\beta r^2}.
\label{eq:19}
\end{align}
Thus this family again belongs to the common radial-Gaussian class,
\begin{align}
P_m(r)\propto r^\nu e^{-\lambda r^2}, \quad \nu=2m+1,\quad \lambda=\beta,
  \label{eq:20}
\end{align}
now on the odd branch. As in Section 3, the branch selection is fixed by geometry: $|\psi_{0m}|^2\sim r^{2m}$, while the two-dimensional radial measure contributes the extra factor $r$.

Using the moment formula (\ref{eq:4}) of Section 2, one finds
\begin{align*}
\langle r\rangle_m = \beta^{-1/2} \frac{\Gamma\!\left(m+\frac32\right)}{\Gamma(m+1)}, \; \langle r^{-1}\rangle_m = \beta^{1/2} \frac{\Gamma\!\left(m+\frac12\right)}{\Gamma(m+1)}\;,
\end{align*}
and therefore
\begin{align}
Q_m^{(\mathrm{FD})} := \langle r\rangle_m\langle r^{-1}\rangle_m = \frac{\Gamma\!\left(m+\frac32\right)\Gamma\!\left(m+\frac12\right)} {\Gamma(m+1)^2}.
\label{eq:21}
\end{align}
With the finite Wallis partial product Eq.~(\ref{eq:11}), the half-integer Gamma identities give
\begin{align}
W_m=\frac{\pi}{2}\,\bigl(Q_m^{(\mathrm{FD})}\bigr)^{-1}.
\label{eq:22}
\end{align}

The reason for this inversion is purely structural. In the oscillator branch, the three-dimensional radial measure contributes an extra factor $r^2$, so that the density takes the form $r^{2\ell+2}e^{-\alpha r^2}$ and the corresponding Gamma ratio places the half-integer factors in the denominator of $Q_\ell^{(\mathrm{osc})}$. In the planar branch, by contrast, the two-dimensional measure contributes only one extra factor $r$, so that the density becomes $r^{2m+1}e^{-\beta r^2}$ and the half-integer Gamma factors appear in the numerator of $Q_m^{(\mathrm{FD})}$. It is this reversal that makes the finite Wallis product arise through $Q^{-1}$ rather than through $Q$ itself.

The semiclassical picture is the planar analogue of the thin-shell localization of the oscillator circular states. The radial density reaches its maximum at
\begin{align}
r_*^2=\frac{2m+1}{2\beta},
\end{align}
and expansion of $\log P_m(r)$ around $r=r_*$ gives
\begin{align}
P_m(r)\approx P_m(r_*)\exp\!\left[-\frac{(r-r_*)^2}{2\sigma_r^2}\right], \; \sigma_r=\frac{1}{2\sqrt{\beta}}.
\end{align}
Hence
\begin{align}
\frac{\sigma_r}{r_*} = \frac{1}{\sqrt{4m+2}} \sim \frac{1}{2\sqrt{m}}, \qquad m\to\infty.
\end{align}

\begin{figure}[!h]
  \centering
  \includegraphics[width=8.7cm]{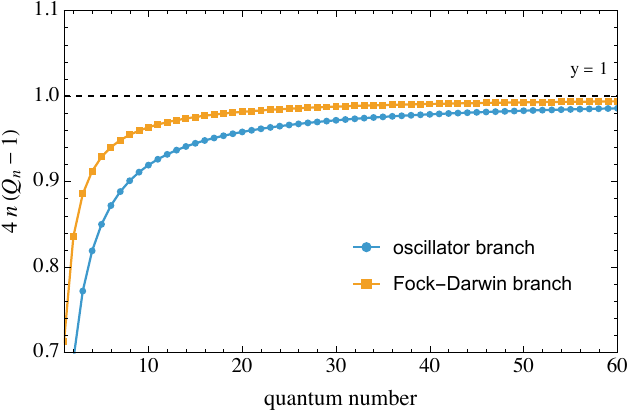}
	\caption{Scaled defects $4\ell(Q_{\ell}^{(\mathrm{osc})}-1)$ and $4m(Q_{m}^{(\mathrm{FD})}-1)$. Both approach unity, showing the common leading asymptotic law $Q=1+\frac{1}{4n}+O(n^{-2})$ underlying the two Wallis-product realizations.}
	\label{fig:Qcompare}
\end{figure}

Thus the large-$m$ states form a narrow annulus of increasing radius and asymptotically vanishing relative thickness. In the lowest-Landau-level interpretation, this is the corresponding guiding-center picture in the correspondence-principle regime.
Accordingly,
\begin{align}
Q_m^{(\mathrm{FD})}
=
1+\frac{1}{4m}-\frac{3}{32m^2}+O(m^{-3}),
\quad m\to\infty,
\end{align}
by the standard asymptotic expansion of Gamma-function ratios. Therefore
\begin{align}
W_m
=
\frac{\pi}{2}\,\bigl(Q_m^{(\mathrm{FD})}\bigr)^{-1}
\longrightarrow
\frac{\pi}{2},
\qquad m\to\infty.
\end{align}
The common large-quantum-number behavior of the oscillator and planar branches is illustrated in Figure~\ref{fig:Qcompare}.

In particular, the figure shows that both branches share the same leading scaled defect, confirming the common first semiclassical correction to reciprocal radial rigidity. Thus the Wallis formula is recovered here because the lowest-radial-branch planar states furnish an exact finite-product realization whose reciprocal-radial observable again approaches its classical limit in the semiclassical regime.

\section{Conclusion}

We have presented a unified quantum-mechanical account of the Wallis formula in two canonical solvable systems: the circular states of the three-dimensional isotropic harmonic oscillator and the lowest-radial-branch states of the planar Fock--Darwin problem, including the lowest Landau level sector. The point of the comparison is not to accumulate separate derivations, but to identify a common radial-Gaussian reciprocal structure and exhibit its two clearest semiclassical realizations.

In both models, the radial probability density belongs exactly to the family $P(r)\propto r^\nu e^{-\lambda r^2}$,so the dimensionless reciprocal observable $Q=\langle r\rangle\langle r^{-1}\rangle$ can be computed in closed form and is independent of the scale parameter. In each case, the relevant Gamma-function ratio lies on a half-integer branch and therefore determines a finite Wallis partial product. The difference between the two realizations is equally simple: the oscillator selects the even branch, for which the partial product is associated with $Q$, whereas the planar magnetic realization selects the odd branch, for which it is associated with $Q^{-1}$.

The physical interpretation is parallel in the two cases. In the oscillator family, the large-$\ell$ states form a thin spherical shell centered asymptotically on the classical circular orbit of the effective radial potential. In the planar magnetic family, the large-$m$ states form a narrow annulus with a natural guiding-center interpretation. In both settings the absolute radial width remains finite while the relative width vanishes, so that $Q\to 1$. It is this asymptotic reciprocal radial rigidity that promotes the exact finite-product identities to the Wallis formula.

In contrast with the earlier hydrogen-based variational route and its later extensions, the present formulation identifies a common exact reciprocal-radial structure realized directly in two canonical solvable branches. This also clarifies why the Wallis product is of interest in quantum mechanics: a classical analytic identity emerges from a simple and physically transparent semiclassical radial structure. These two models thus appear not as isolated examples, but as canonical realizations of a broader Wallis-bearing structural pattern in solvable radial quantum families, one that reappears whenever an exact reciprocal-radial observable is governed by a half-integer Gamma-function channel.


\appendix
\section{Exact finite Wallis forms}

For the oscillator branch,
\begin{align*}
Q_\ell^{(\mathrm{osc})}
=
\frac{\Gamma(\ell+1)\Gamma(\ell+2)}
{\Gamma\!\left(\ell+\frac32\right)^2}.
\end{align*}
Using
\begin{align*}
\Gamma(\ell+1)=\ell!,\;&
\Gamma(\ell+2)=(\ell+1)!,\\
\Gamma\!\left(\ell+\frac32\right)
=&
\frac{(2\ell+1)!!}{2^{\ell+1}}\sqrt{\pi},
\end{align*}
one obtains
\begin{align*}
Q_\ell^{(\mathrm{osc})}
=&
\frac{2}{\pi}
\frac{(2\ell)!!(2\ell+2)!!}{[(2\ell+1)!!]^2} \\
=&
\frac{2}{\pi}
\prod_{j=1}^{\ell+1}\frac{(2j)^2}{(2j-1)(2j+1)}.
\end{align*}
Hence
\[
W_{\ell+1}=\frac{\pi}{2}\,Q_\ell^{(\mathrm{osc})}.
\]

For the planar branch,
\[
Q_m^{(\mathrm{FD})}
=
\frac{\Gamma\!\left(m+\frac32\right)\Gamma\!\left(m+\frac12\right)}
{\Gamma(m+1)^2}.
\]
Inverting and using
\begin{align*}
\Gamma\!\left(m+\frac12\right)
=&
\frac{(2m)!}{4^m m!}\sqrt{\pi},
\\
\Gamma\!\left(m+\frac32\right)
=&
\left(m+\frac12\right)\Gamma\!\left(m+\frac12\right),
\end{align*}
one finds
\[
\frac{\pi}{2}\,\bigl(Q_m^{(\mathrm{FD})}\bigr)^{-1}
=
\prod_{k=1}^{m}\frac{(2k)^2}{(2k-1)(2k+1)}
=
W_m.
\]


\vspace{1cm}
\bibliographystyle{JHEP}

\bibliography{WallisReferences}

\end{document}